\documentclass[superscriptaddress,amsmath,amssymb,nofootinbib,eqsecnum,a4paper,secnumarabic,11pt]{revtex4}         
\usepackage[pdftex]{graphicx}
\usepackage{hyperref}     
\usepackage{grffile} 
\usepackage{slashed,ulem,color}

\newcommand{\VEV}[1]{\left\langle #1 \right\rangle}

\usepackage{titlesec}     
\newcommand*{\justifyheading}{\raggedright}
\titleformat{\chapter}[display]
  {\normalfont\huge\bfseries\justifyheading}{\chaptertitlename\ \thechapter}
  {20pt}{\Huge}
\titleformat{\section}
  {\normalfont\Large\bfseries\justifyheading}{\thesection}{1em}{}
\titleformat{\subsection}
  {\normalfont\large\bfseries\justifyheading}{\thesubsection}{1em}{}
\titleformat{\subsubsection}
  {\normalfont\bfseries\justifyheading}{\thesubsubsection}{1em}{}
\usepackage[english]{babel}

\begin{document}

\title{Early kinetic decoupling and a pseudo-Nambu-Goldstone dark matter model}
\author{Tomohiro Abe}
\affiliation{
  Department of Physics, Faculty of Science and Technology, Tokyo University of Science,
  Noda, Chiba 278-8510, Japan
}
\email{abe.tomohiro@rs.tus.ac.jp}

\begin{abstract}
We study the early kinetic decoupling effect in a pseudo-Nambu-Goldstone (pNG) dark matter (DM) model.
The pNG DM scattering processes with particles in the thermal bath in the early Universe are suppressed by the small momentum transfer.
As a result, kinetic equilibrium is not maintained, and the temperature of DM is different from the temperature of the thermal bath at the freeze-out era.
This temperature difference affects the thermal relic abundance of DM.
We investigate the early kinetic decoupling in the Higgs resonance region, 50~GeV $\lesssim m_\chi \lesssim m_h/2$, where $m_\chi$ is the mass of the DM, and $m_h/2 \simeq$ 62.5~GeV. 
We find that the DM-Higgs coupling determined to obtain the measured value of the DM energy density is underestimated in the literature. The enhancement in the coupling leads larger value of the Higgs invisible decay rate. It enlarges the capability to discover the DM signals from the decay of the Higgs bosons at collider experiments.
\end{abstract}

\maketitle


\section{Introduction}

A weakly interacting massive particle (WIMP) is a well-known dark matter (DM) candidate.
It couples to the standard model (SM) particles weakly, and pairs of DM particles annihilate into the SM particles.
If the annihilation cross section is $\sim {\cal O}(10^{-26})$ cm$^3 $s$^{-1}$, 
then the measured value of the DM energy density by the Plank Collaboration~\cite{1807.06209} is easily explained by the freeze-out mechanism~\cite{Lee:1977ua}.
The interaction between WIMPs and SM particles typically predicts WIMP-nucleon scattering processes as well. On the other hand, direct detection experiments give stringent upper bound on the DM-nucleon scattering cross section~\cite{1608.07648, 1708.06917, 1805.12562}. 
Therefore, the DM-SM scattering processes must be suppressed while keeping the DM annihilation processes into the SM particles.

Resonance enhancement in the annihilation processes 
is utilized to suppress the scattering cross section while keeping the annihilation cross section. 
It requires a small DM-mediator coupling to obtain the right amount of the DM relic abundance by the freeze-out mechanism, and thus the scattering cross section becomes small by the small coupling.
Another way to suppress the scattering processes is to rely on models that predict scattering amplitudes suppressed by the low momentum transfer~\cite{1203.2064, 1404.3716, 1609.09079, 1612.06462, 1708.02253}.

The suppressed scattering processes of DM and SM particles can make the kinetic decoupling happen earlier. Suppose the dark sector and the visible sector are in the kinetic equilibrium. In that case, the temperature of the dark matter $(T_\chi)$ is the same as the temperature of the visible sector $(T)$. The kinetic equilibrium is maintained as long as the scattering processes between the DM and the visible particles are efficient compared to the expansion rate of the Universe. 
In most WIMP models, the kinetic equilibrium is maintained during the chemical decoupling. It allows us to calculate the thermal relic abundance of DM under the assumption that $T_\chi = T$.
However, if the scattering processes are suppressed, it is unclear whether the kinetic equilibrium is maintained during the chemical decoupling era. 
In that case, we have to calculate the time evolution of $T_\chi$ as well as the DM number density by solving the Boltzmann equation. 
It was shown that the kinetic decoupling happens earlier than usual
in a scalar singlet DM model~\cite{1706.07433, 1912.02870},
a scalar $Z_3$ singlet DM model~\cite{1901.08074},
and a fermionic DM model~\cite{2004.10041}.

In this paper, we focus on the pseudo-Nambu-Goldstone (pNG) DM model~\cite{1708.02253}
and study the DM-Higgs coupling in the Higgs resonance region with the effect of the early kinetic decoupling.
Here the Higgs resonance region means
50~GeV $\lesssim m_\chi \lesssim m_h/2$, where $m_\chi$ is the mass of the DM, and $m_h/2 \simeq$ 62.5~GeV. 
We assume the coupling is determined by the freeze-out mechanism.
The same analysis in other models~\cite{1706.07433, 1901.08074, 1912.02870, 2004.10041} shows the enhancement in DM-Higgs coupling compared to 
the analysis under the assumption that $T_\chi = T$.
A significant enhancement in the DM-Higgs coupling is expected in the pNG model because it predicts highly suppressed scattering processes by low momentum transfer,

The rest of this paper is organized as follows.
In Sec.~\ref{sec:Model}, we review the pNG DM model.
In Sec.~\ref{sec:method}, we briefly explain how to calculate the DM number density without assuming $T_\chi = T$.
The result is shown in Sec.~\ref{sec:result}.
It is shown that the DM-Higgs coupling in the Higgs resonance region is underestimated in the literature.
Section~\ref{sec:summary} is devoted for the conclusion.

\section{Model}\label{sec:Model}
We briefly review the pNG model proposed in~\cite{1708.02253}.
A complex gauge singlet field $S$ is introduced into the SM.
The fermion and gauge sectors of the model are the same as in the SM.
The Lagrangian that contains $S$ is given by 
\begin{align}
\left.
 {\cal L}
\right|_{scalar}
= 
  D^\mu H^\dagger D_\mu H
+ \partial^\mu S^* \partial_\mu S
-V_\text{scalar}
-V_\text{soft}
,
\end{align}
where
\begin{align}
V_\text{scalar}
=&
- \frac{\mu_H^2}{2} H^\dagger H  
+ \frac{\lambda_H}{2} \left( H^\dagger H \right)^2
- \frac{\mu_s^2}{2} S^* S
+ \frac{\lambda_s}{2} (S^* S)^2
+ \lambda_{hs} H^\dagger H S^* S  
,\\
V_\text{soft}
=&
- \frac{\mu_s^{\prime 2}}{4} \left( S^2 + S^{* 2}\right) 
\label{eq:Vsoft}
,
\end{align}
and $H$ is the SM Higgs field.
The Lagrangian has a global U(1) symmetry that rotates only $S$ as $S \to e^{i\alpha} S$
in the kinetic term and $V_{scalar}$. 
This symmetry is explicitly broken in $V_{soft}$, 
but the whole Lagrangian is still invariant under a $Z_2$ symmetry, $S \to S^*$.
If $\mu_s'^2 =0$, then the U(1) global symmetry would be exact, and a Nambu-Goldstone (NG) boson arises after $S$ develops its vacuum expectation value (VEV). 
We can take the VEV of $S$ as real without a loss of generality, so the imaginary part of $S$ is regarded as the NG boson.
Since the U(1) is explicitly broken, the NG boson obtains mass and becomes a pNG boson.
This pNG boson is odd under the $Z_2$ symmetry ($S \to S^*$), and all the other particles are even.
Hence this $Z_2$ stabilizes the pNG and makes it as the DM candidate in this model.

We can introduce other U(1) breaking terms such as $S + S^*$.
However, the U(1) breaking terms other than in $V_{soft}$ break the desired property
that suppresses the DM scattering processes by momentum transfer.\footnote{
Even in that case, the scattering process can be suppressed if we assume
the degenerated mass spectra~\cite{2101.04887}.
} 
UV complete models that forbid the U(1) breaking terms other than $V_{soft}$ are
proposed in~\cite{2001.03954,2001.05910, 2104.13523, 2105.03419}.

Component fields and VEVs of the singlet and the SM Higgs fields are parametrized as
\begin{align}
 S = \frac{v_s + s + i \chi }{\sqrt{2}}, 
\quad
 H = \begin{pmatrix}
      i \pi_{W^+} \\  \frac{v + \sigma - i \pi_Z}{\sqrt{2}}
     \end{pmatrix}
,
\end{align}
where $v_s$ and $v$ are the VEVs, 
$\chi$ is the DM, $s$ and $\sigma$ are CP-even scalar bosons, and
$\pi_{W^+}$ and $\pi_Z$ are the would-be NG bosons for $W^+$ and $Z$, respectively.
The stationary condition of this vacuum imposes the following relations for the mass parameters:
\begin{align}
 \mu_H^2 =& v^2 \lambda_H + v_s^2 \lambda_{hs} , \\ 
 \mu_s^2 =& -\mu_s^{\prime 2} + v^2 \lambda_{hs}  + v_s^2 \lambda_s. 
\end{align}
The mass terms of the physical scalar particles are given by
\begin{align}
 {\cal L}_\text{mass}^\text{scalar}
=&
- \frac{1}{2} \mu_s^{\prime 2} \chi^2
- \frac{1}{2} 
 \begin{pmatrix}  \sigma & s \end{pmatrix}
 \begin{pmatrix}
  \lambda_H v^2  &  \lambda_{hs} v v_s \\
  \lambda_{hs} v v_s  &  \lambda_s v_s^2 
 \end{pmatrix}
 \begin{pmatrix}  \sigma \\ s \end{pmatrix}
.
\end{align}
The mass eigenstates that are denoted by $h$ and $h'$ are obtained by diagonalizing the
two-by-two mass matrix above.
The relation between the mass eigenstates and component fields are given by 
\begin{align}
  \begin{pmatrix}  \sigma \\ s \end{pmatrix}
=&
\begin{pmatrix}
  c_{h}  & s_{h} \\
 -s_{h}  & c_{h} 
\end{pmatrix}
 \begin{pmatrix}  h \\ h' \end{pmatrix}
,
\end{align}
where
 $c_{h} = \cos\theta_h$
and 
 $s_{h} = \sin\theta_h$.

There are six model parameters in the scalar sector,
\begin{align}
\left(
\mu_H^2,\ \mu_s^2,\   \lambda_H,\ \lambda_s,\ \lambda_{hs},\
\mu_s^{\prime 2}
\right).
\end{align}
Instead of using these parameters, we choose the following six parameters as inputs in the following analysis:
\begin{align}
\left( v,\ v_s,\   m_h,\ m_{h'},\ \theta_{h},\ m_\chi \right),
\end{align}
where $m_X$ is the mass of $X$.
Among these parameters, $v$ and $m_h$ are already known, $v \simeq 246$~GeV and $m_h \simeq$ 125~GeV,
and thus we have four free parameters in the following analysis.

In the following, we discuss the annihilation processes of pairs of DM particles into the SM fermions
and DM-SM fermion elastic scattering processes.
These processes are mediated by $h$ and $h'$, and the following interaction terms are essential:
\begin{align}
 {\cal L}\supset&
 - \frac{1}{2} g_{\chi \chi h} \chi^2 h
 - \frac{1}{2} g_{\chi \chi h'} \chi^2 h'
 - g_{\bar{f} f h} \bar{f} f h 
 - g_{\bar{f} f h'} \bar{f} f h' 
,
\end{align}
where
\begin{align}
 g_{\chi \chi h}
=& - \frac{m_h^2}{v_s} s_{h} 
\label{eq:g_chichih}
,\\
 g_{\chi \chi h'}
=& + \frac{m_{h'}^2}{v_s} c_{h} 
 ,\\
 g_{\bar{f} f h}
=& + \frac{m_{f}}{v} c_{h} 
 ,\\
 g_{\bar{f} f h'}
=& + \frac{m_{f}}{v} s_{h}.
\end{align}
Here, $f$ stands for the SM fermions.

An important property of the pNG DM model is that
the scattering processes are suppressed by the momentum transfer.
It is easy to calculate the DM-SM fermion scattering amplitude at the tree level.
The square of the scattering amplitude for $\chi f \to \chi f$ is given by
\begin{align}
{\cal S}_f \equiv \sum_{d.o.f} |{\cal M}_{\chi f \to \chi f}|^2
=&
2 \frac{m_f^2}{v^2} 
\left(
\frac{g_{\chi \chi h} c_h }{t-m_h^2} 
+
\frac{g_{\chi \chi h'} s_h }{t-m_{h'}^2} 
\right)^2
(4 m_f^2 - t)
N_c
\nonumber\\
=&
2 m_f^2
N_c
\frac{v^2}{v_s^2}
s_h^2 c_h^2
\frac{(m_h^2 - m_{h'}^2)^2}{v^4}
\frac{t^2 (4 m_f^2 - t)}{(t-m_h^2)^2 (t-m_{h'}^2)^2}
,
\label{eq:scatt-amp}
\end{align}
where $N_c = 3 \ (1)$ for quarks (leptons).
The summation is taken for all the internal degrees of freedom for all the initial and final states.
The processes for $\chi \bar{f} \to \chi \bar{f}$ result in the same.
As can be seen, 
${\cal S}_f$ is proportional to the fourth power of the momentum transfer, or $t^2$.
The momentum transfer is small in the scattering process at the direct detection experiments,
and thus the model evades the constraint on the spin-independent WIMP-nucleon scattering cross section.
The momentum transfer is also small in the early Universe after DM becomes nonrelativistic.
Thus the kinetic decoupling happens earlier than usual as we will see below.

Before closing the section, we discuss the DM-DM scattering for later convenience.
In this model, the DM-DM scattering is generated by the $h$ and $h'$ exchanging diagrams and 
the contact interaction terms due to the $\lambda_s$ term given by
\begin{align}
 {\cal L} \supset& 
- \frac{\lambda_s}{8} \chi^4
= - \frac{3 m_h^2 s_h^2 + 3 m_{h'}^2 c_h^2}{8 v_s^2} \chi^4
.
\end{align}
We find the invariant amplitude of $\chi \chi \to \chi \chi$ at the tree level as
\begin{align}
 i {\cal M}_{\chi \chi \to \chi \chi}
=&
-i \frac{m_h^2 s_h^2}{v_s^2} 
\left( \frac{s}{s-m_h^2} + \frac{t}{t-m_h^2} + \frac{u}{u-m_h^2} \right)
-i \frac{m_{h'}^2 c_h^2}{v_s^2} 
\left( \frac{s}{s-m_{h'}^2} + \frac{t}{t-m_{h'}^2} + \frac{u}{u-m_{h'}^2} \right)
.
\label{eq:DM-self}
\end{align}
In the nonrelativistic limit, $s \simeq 4 m_\chi^2$, $t \simeq u \simeq 0$, and thus
\begin{align}
 {\cal M}_{\chi \chi \to \chi \chi}
\simeq&
- \frac{m_h^2 s_h^2}{v_s^2} \frac{m_\chi^2}{m_\chi^2-m_h^2} 
- \frac{m_{h'}^2 c_h^2}{v_s^2} \frac{m_\chi^2}{m_\chi^2-m_{h'}^2} 
.
\label{eq:DM-self-nonrela}
\end{align}
This amplitude is not suppressed.
By utilizing this process, DM particles make a thermal bath in the dark sector, and the temperature of the DM can be defined.

\section{Method}\label{sec:method}

We briefly describe how to calculate the DM number density with 
the effect of the early kinetic decoupling based on Ref.~\cite{1706.07433}.\footnote{
The authors of Ref.~\cite{1706.07433} recently developed a public code to 
obtain the relic abundance with the early kinetic decoupling effect~~\cite{2103.01944}.
}

The Boltzmann equation for our Universe is given by
\begin{align}
 E 
\left( \frac{\partial}{\partial t} - H \vec{p}\cdot \frac{\partial}{\partial \vec{p}} \right)
f_\chi (t, \vec{p})
=
C_{ann.}[f_\chi] + C_{el.}[f_\chi]
,
\label{eq:boltzmann-eq}
\end{align}
where $E$ is the energy of the DM, $H$ is the Hubble constant, $\vec{p}$ is the momentum of DM,
and $f_\chi$ is the phase-space density of DM.
The collision term is divided into two parts. 
One is for the annihilation of pairs of DM particles ($C_{ann.}$), 
and the other is for elastic scatterings of a DM particle off an SM particle in the thermal bath ($C_{el.}$).
For two-to-two processes, they are given by
\begin{alignat}{3}
 C_{ann.}
=&
\frac{1}{2 g_\chi}
\sum_{d.o.f}
&
&
\int \frac{d^3 p'}{(2\pi)^3 2 E_{p'}}
\int \frac{d^3 k}{(2\pi)^3 2 E_k}
\int \frac{d^3 k'}{(2\pi)^3 2 E_{k'}}
(2\pi)^4 \delta^4(p + p' - k - k')
\nonumber\\
&&& 
\times \Bigl(
-|{\cal M}_{\chi \chi \to {\cal B} {\cal B}'}|^2 f_\chi(\vec{p}) f_\chi(\vec{p'}) (1 \pm f^{eq}_{\cal B}(\vec{k})) (1 \pm f^{eq}_{\cal B'}(\vec{k'}))
\nonumber\\
&&& \qquad
+|{\cal M}_{{\cal B}{\cal B'} \to \chi \chi}|^2 f^{eq}_{\cal B}(\vec{k}) f^{eq}_{{\cal B}'}(\vec{k'}) (1 \pm f_\chi(\vec{p})) (1 \pm f_\chi(\vec{p'}))
\Bigr)
,\\
 C_{el.}
=&
\frac{1}{2g_\chi}
\sum_{d.o.f}
&
&
\int \frac{d^3 p'}{(2\pi)^3 2 E_{p'}}
\int \frac{d^3 k}{(2\pi)^3 2 E_k}
\int \frac{d^3 k'}{(2\pi)^3 2 E_{k'}}
(2\pi)^4 \delta^4(p + p' - k - k')
\nonumber\\
&&&
\times \Bigl(
-|{\cal M}_{\chi {\cal B} \to \chi {\cal B}}|^2 
f_\chi(\vec{p}) f^{eq}_{\cal B}(\vec{k}) (1 \pm f_\chi(\vec{p'})) (1 \pm f^{eq.}_{\cal B}(\vec{k'}))
\nonumber\\
&&& \qquad 
+|{\cal M}_{\chi {\cal B} \to \chi {\cal B}}|^2 
f_\chi(\vec{p'}) f^{eq.}_{\cal B}(\vec{k'}) (1 \pm f_\chi(\vec{p})) (1 \pm f^{eq.}_{\cal B}(\vec{k}))
\Bigr)
,
\label{eq:c_el_body}
\end{alignat}
where 
${\cal B}$ and ${\cal B}'$ stand for particles in the thermal bath such as quarks, 
$g_\chi$ is the number of internal degrees of freedom of DM,
and
 $f_{\cal B}^{eq}$ is given by the Fermi-Dirac or Bose-Einstein distribution depending on the spin of ${\cal B}$.
The summation should be taken for all the internal degrees of freedom
for all the particles.

In the following analysis, we assume that the DM is in the thermal bath in the dark sector.
This assumption is justified if DM-DM elastic scattering processes exist.
Although the DM scattering off the particles in the thermal bath is suppressed by the 
small momentum transfer in the pNG model (see Eq.~\eqref{eq:scatt-amp}), 
the DM-DM scattering process is not suppressed kinematically as shown in Eqs.~\eqref{eq:DM-self} and \eqref{eq:DM-self-nonrela}.
Therefore, this assumption is justified,
and we can safely introduce the temperature of DM ($T_\chi$).
As a result, $f_\chi$ is given by
\begin{align}
 f_\chi =& \alpha(T_\chi) e^{- \frac{E_\chi}{T_\chi}}
= \frac{n_\chi}{n_\chi^{eq.}} e^{- \frac{E_\chi}{T_\chi}},
\label{eq:fchi}
\end{align}
where $n_\chi$ is the number density of DM defined by
\begin{align}
 n_\chi
=&
g_\chi
\int \frac{d^3 p}{(2\pi)^3} f_\chi (\vec{p}),
\label{def:n_chi}
\end{align}
 and 
$n_\chi^{eq.}$ is the same but for $T = T_\chi$, namely,
\begin{align}
 n_\chi^{eq}(T_\chi) 
=& 
 \int \frac{d^3 p}{(2\pi)^3} e^{- \frac{E_p}{T_\chi}}
=
 \frac{m_\chi^2 T_\chi}{2 \pi^2} K_2\left( \frac{m_\chi}{T_\chi} \right)
,
\end{align}
where $K_2$ is the modified Bessel function of the second kind.
The relation between the temperature and the distribution function of the DM is given by
\begin{align}
 T_\chi
=&
\frac{g_\chi}{n_\chi}
\int \frac{d^3 p}{(2\pi)^3} \frac{\vec{p}^2}{3 E} f_\chi (\vec{p})
.
\label{def:T_chi}
\end{align}
Using this $f_\chi$ given in Eq.~\eqref{eq:fchi},
we simplify ${\cal C}_{el.}$ given in Eq.~\eqref{eq:c_el_body}.
The detail is discussed in the Appendix.

Using Eqs.~\eqref{eq:boltzmann-eq}, \eqref{def:n_chi}, and \eqref{def:T_chi},
we obtain the coupled differential equations, which describe the time evolution of the number density and temperature of DM without assuming $T_\chi = T$.
Instead of using $n_\chi$ and $T_\chi$,
we use $Y$ and $y$ defined by
\begin{align}
 Y =& \frac{n_\chi}{s},\\
 y =& \frac{m_\chi T_\chi}{s^{2/3}},
\end{align}
where $s$ is the entropy density, which is a function of $T$,
\begin{align}
  s =& \frac{2 \pi^2}{45} g_s(T) T^3,
\end{align}
where $g_s$ is the effective degrees of freedom for the entropy density.
We also introduce dimensionless parameters given by
\begin{align}
 x =& \frac{m_\chi}{T},\\
 x_\chi =& \frac{m_\chi}{T_\chi}.
\end{align}
Finally, we find the following coupled differential equations:
\begin{align}
 \frac{dY}{dx}
=&
\sqrt{\frac{8 m_{pl}^2 \pi^2}{45}} \frac{m_\chi}{x^2} \sqrt{g_* (T)}
\left(
-\VEV{\sigma v}_{T_\chi} Y^2
+\VEV{\sigma v}_{T} Y_{eq}^2
\right)
\label{eq:dYdx},\\
\frac{1}{y} \frac{dy}{dx}
=&
\sqrt{\frac{8 m_{pl}^2 \pi^2}{45}} \frac{m_\chi}{x^2} \sqrt{g_* (T)}
\Biggl\{
Y \left(
\VEV{\sigma v}_{T_\chi} 
-\VEV{\sigma v}_{2, T_\chi}
\right)
+
\frac{Y_{eq}^2}{Y} \left(
\frac{y_{eq}}{y}
\VEV{\sigma v}_{2, T} 
-\VEV{\sigma v}_{T}
\right)
\Biggr\}
\nonumber\\
&
+
\frac{\sqrt{g_* (T)}}{g_s(T)}
\tilde{\delta}
+
\left(
1 + \frac{T}{3 g_s(T)} \frac{d g_s(T)}{dT}
\right)
\frac{1}{3 m_\chi} \frac{y_{eq}}{y} 
\VEV{\frac{p^4}{E^3}}
,
\label{eq:dydx}
\end{align}  
where
\begin{align}
 \sqrt{g_*(T)}
=&
 \frac{g_s(T)}{\sqrt{g(T)}} \left( 1 + \frac{T}{3 g_s(T)} \frac{d g_s(T)}{dT} \right)
,\\
\VEV{\sigma v}_{T_\chi}
=&
 \frac{x_\chi}{8 m_\chi^5 \left[K_2\left( x_\chi \right)\right]^2}
\int_{4 m_\chi^2}^{\infty}
 ds
 K_1\left( \frac{\sqrt{s}}{T_\chi} \right)
 s \sqrt{1 - \frac{4 m_\chi^2}{s}}
\tilde{f}(s)
,\\
\VEV{\sigma v}_{2, T_\chi}
=&
\frac{x_\chi^3}{48 m_\chi^6 \left[K_2\left( x_\chi \right)  \right]^2}
\int_{4 m_\chi^2}^\infty ds 
s^{3/2} 
\tilde{f}(s)
\tilde{g}\left(\frac{s}{4m_\chi^2}\right)
,\\
\tilde{\delta}
=&
 - \frac{15}{128\pi^5} 
\sqrt{\frac{8 m_{pl}^2 \pi^2}{45}} 
 \frac{m_\chi}{T_\chi^2}
\frac{e^{-x_\chi}}{K_2(x_\chi)}
\int_0^1 d \omega
\int_0^1 dy
\int_0^1 dz
\nonumber\\
& \qquad \qquad \qquad \times
\sqrt{\epsilon_1^2 - 1}
\sqrt{\epsilon_2^2 - 1}
\epsilon_1^2 ( \epsilon_1 - 1)
(\epsilon_1 - \epsilon_2)
\left(
1 + \frac{1}{\epsilon_1 \epsilon_2}
\right)
\nonumber\\
& \qquad \qquad \qquad \times
\frac{-e^{-(\epsilon_1 -1)x_\chi} + e^{(\epsilon_2 - \epsilon_1) x} e^{- (\epsilon_2 -1)x_\chi}}{ 1 - e^{x(\epsilon_2 - \epsilon_1)}}
\frac{1}{\sqrt{(\epsilon_1 - \epsilon_2)^2 - \frac{t}{m_\chi^2}}}
\sum_{f}{\cal S}_f
\nonumber\\
& 
\qquad \qquad \qquad \times
\ln
\left(
\frac{1 + e^{-\frac{x(\epsilon_1 - \epsilon_2)}{2}} \exp\left(-\frac{x}{2}\sqrt{(\epsilon_1 - \epsilon_2)^2 - \frac{t}{m_\chi^2}}\sqrt{1 - \frac{4 m_f^2}{t}}\right)
}{
1 + e^{\frac{x(\epsilon_1 - \epsilon_2)}{2}}\exp\left(-\frac{x}{2}\sqrt{(\epsilon_1 - \epsilon_2)^2 - \frac{t}{m_\chi^2}}\sqrt{1 - \frac{4 m_f^2}{t}}\right)
}
\right)
,\\
\VEV{\frac{p^4}{E^3}}
=&
\frac{1}{n_\chi^{eq}(T_\chi)}
\int \frac{d^3 p}{(2\pi)^3} \frac{\left( \vec{p} \cdot \vec{p} \right)^2}{E^3} e^{- \frac{E}{T_\chi}}
,
\end{align}
$m_{pl}$ is the reduced Plank mass, $m_{pl} = 1.220910 \times 10^{19} (8\pi)^{-1/2}$~GeV,
$g$ is the effective degrees of freedom for the energy density,
$K_1$ is the modified Bessel function of the first kind,
${\cal S}_f$ is defined in Eq.~\eqref{eq:scatt-amp}, and
\begin{align}
 \Gamma_h(\sqrt{s})
=&
 c_{\theta_h}^2 \Gamma_h^\text{SM}(\sqrt{s})
+ \frac{1}{32 \pi} \frac{g_{\chi\chi h}^2}{\sqrt{s}} \sqrt{1 - \frac{4 m_\chi^2}{s}}
 \theta(\sqrt{s} - 2 m_\chi)
,\\
 \Gamma_{h'}(\sqrt{s})
=&
 s_{\theta_h}^2 \Gamma_h^\text{SM}(\sqrt{s})
+ \frac{1}{32 \pi} \frac{g_{\chi\chi h'}^2}{\sqrt{s}} \sqrt{1 - \frac{4 m_\chi^2}{s}}
 \theta(\sqrt{s} - 2 m_\chi)
,\\
\tilde{f}(s)
=&
\left|
\frac{g_{\chi \chi h} c_{\theta_h}}{s - m_h^2 + i \sqrt{s} \Gamma_h(\sqrt{s})}
+
\frac{g_{\chi \chi h'} s_{\theta_h}}{s - m_{h'}^2 + i \sqrt{s} \Gamma_{h'}(\sqrt{s})}
\right|^2
\Gamma^{SM}_{h}(\sqrt{s})
,\\
\tilde{g}(\tilde{s})
=&
\int_1^{\infty} d\epsilon_+ \exp\left( - 2 \frac{m_\chi}{T_\chi} \sqrt{\tilde{s}} \epsilon_+ \right)
\left\{
2 \epsilon_+ \sqrt{ (\tilde{s} - 1)(\epsilon_+^2 -1)  }
+ \frac{1}{\sqrt{\tilde{s}}} \ln \frac{\epsilon_+ \sqrt{\tilde{s}} - \sqrt{ (\tilde{s} - 1)(\epsilon_+^2 -1) } }{\epsilon_+ \sqrt{\tilde{s}} + \sqrt{ (\tilde{s} - 1)(\epsilon_+^2 -1) } }
\right\}
,\\
 \epsilon_1 =& \frac{1}{1-\omega},\\
 \epsilon_2 =& 1+ \frac{\omega y}{1-\omega},\\
 t =& m_\chi^2 
\left( 4 \sqrt{\epsilon_1^2 - 1} \sqrt{\epsilon_2^2 - 1} z + (\epsilon_1-\epsilon_2)^2 -( \sqrt{\epsilon_1^2 - 1} + \sqrt{\epsilon_2^2 - 1} )^2  \right).
\end{align}
Here $\Gamma_h^\text{SM}(\sqrt{s})$ is the decay width of the Higgs boson into the SM particle. 
We use the table given by the Higgs cross section working group~\cite{1307.1347} to evaluate $\Gamma_h^\text{SM}(\sqrt{s})$.\footnote{
The table is given at \url{https://twiki.cern.ch/twiki/pub/LHCPhysics/CERNYellowReportPageAt8TeV2014/Higgs_XSBR_YR3_update.xlsx}
.}
For 
$\VEV{\sigma v}_{T} $
and 
$\VEV{\sigma v}_{2, T} $,
replace $T_\chi$ by $T$ in 
$\VEV{\sigma v}_{T_\chi} $
and 
$\VEV{\sigma v}_{2, T_\chi} $, respectively.
Note that the summation for the fermions runs over both fermions and antifermions.

During the QCD phase transition, 
we cannot treat particles as free particles.
Dedicated studies are required for that regime.
In Ref.~\cite{1503.03513},
the table is provided for $g_*$ and $g_s$ for $0.036$~MeV $\lesssim T \lesssim 8.6$~TeV.
Since the values of $g_*$ and $g_s$ do not change for $T \lesssim 0.036$~MeV,
we can regard the values of $g_*$ and $g_s$ at $T = 0.036$~MeV as the values at the temperature today.

We solve 
Eqs.~\eqref{eq:dYdx} and \eqref{eq:dydx}
numerically with the following initial condition:
\begin{align}
 Y(x_{ini.}) =& Y_{eq}(x_{ini.}),\\
 y(x_{ini.}) =& y_{eq}(x_{ini.}),
\end{align} 
where $x_{ini.} \simeq 10$.
This initial condition for $Y$ and $y$ is the assumption, 
but it is reasonable if 
DM annihilation and 
DM-SM scattering processes frequently happen at high temperatures.
The result is not sensitive to the choice of $x_{ini.}$ as long as $x_{ini.} \lesssim 20$.
After solving the coupled differential equations and obtain $Y(x_0)$, where $x_0$ is 
defined by the temperature of the current universe $T_0$ as 
$x_0 = m_\chi/T_0$,
we convert $Y(x_0)$ into $\Omega h^2$ that is given by
\begin{align}
 \Omega h^2
= \frac{m_\chi s_0 Y(x_0)}{\rho_{cr.} h^{-2}}
,
\end{align}
where~\cite{Tanabashi:2018oca}
\begin{align}
 s_0 =& \frac{2 \pi^2}{45} g_s(x_0) T_0^3,\\
 \rho_{cr.} h^{-2} =& 1.05371 \times 10^{-5} \text{ [GeV} \text{ cm}^{-3}],\\
 T_0 =& 2.35 \times 10^{-13} \text{ [GeV]}.
\end{align}
The measured value of $\Omega h^2$ by the Planck Collaboration
is $\Omega h^2 = 0.120\pm 0.001$~\cite{1807.06209}.
We can use this value to determine a model parameter.

In the mass range we focus on, the freeze-out happens around $T \simeq {\cal O}(1)$~GeV.
This temperature is not far from the temperature of the QCD phase transition.
Hence the scattering rate of DM and quarks in the thermal bath is potentially affected by the details of the QCD phase transition. 
Following Ref.~\cite{1706.07433}, we investigate the two extreme scenarios, QCD-A and QCD-B.
In the QCD-A scenario, it is assumed that all quarks are free particles and present in the thermal bath down to $T_c = 154$~MeV~\cite{1205.1914}. 
In the QCD-B scenario, only the light quarks ($u$, $d$, $s$) contribute to the scattering above $4 T_c \sim 600$~MeV~\cite{0903.0189}.
The difference between these two scenarios is whether charm and bottom quarks contribute to the elastic scattering processes or not. Since the scattering rate is proportional to the squared of the quark Yukawa couplings and the color factor, the absence of the heavy quarks can make a large difference between these two scenarios. 
The scattering ratio in the QCD-B is smaller than the one in the QCD-A.
However, as we will see below, the difference between the QCD-A and QCD-B is almost negligible in a viable parameter region in this model. This is because the scattering process is already highly suppressed by the low momentum transfer.

\section{Result}\label{sec:result}
We discuss the effect of the early kinetic decoupling on the DM-Higgs coupling.
In particular, we focus on the Higgs resonant region, where 
50~GeV $\lesssim m_\chi \lesssim m_h/2$, 
and show the impact on the Higgs invisible decay branching ratio. 
A reason why we do not focus on $m_\chi \lesssim 50$~GeV is the bound from the Higgs invisible decay~\cite{1708.02253}. We will discuss more details about the constraint from the Higgs invisible decay below.
Another reason is that the early kinetic decoupling effect is weakened if $m_\chi$ is away from the Higgs resonant region as we will see below. 
The DM-Higgs coupling is determined so as to obtain the measured value of the DM energy density
by solving the coupled differential equations given in 
Eqs.~\eqref{eq:dYdx} and \eqref{eq:dydx}.
The suppression by the low momentum transfer in the scattering processes is strong.
Hence, it is expected that 
the DM-Higgs couplings obtained with and without assuming $T_\chi = T$ are very different.
This difference affects to the prediction of the branching ratio of the SM Higgs boson into a pair of DM particles because it is proportional to the DM-Higgs coupling squared.

\begin{figure}[tb]
\includegraphics[width=0.68\hsize]{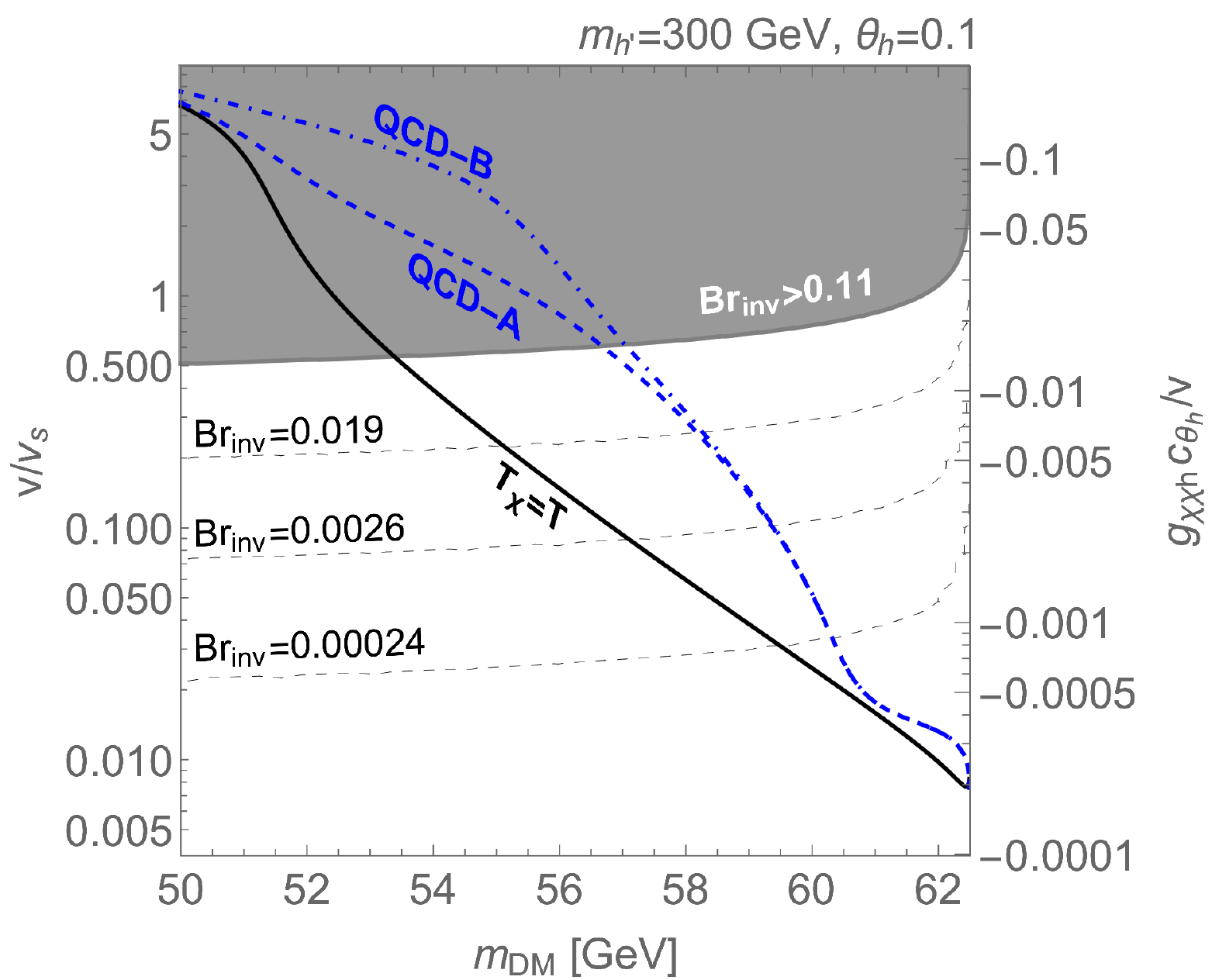}
\caption{
The values of $v/v_s$ that can explain the measured value of the DM energy density, $\Omega h^2 = 0.12$, for a given parameter set.
The values of $g_{\chi \chi h} c_{\theta_h}/v$ are also shown.
We take $m_{h'} = 300$~GeV and $\theta_h = 0.1$.
The black-solid curve is the result from the assumption $T_\chi = T$, namely the standard treatment in WIMP calculation. The blue dashed and blue dot-dashed curves are for the QCD-A and QCD-B, respectively. The gray shaded region is already excluded by the Higgs invisible decay at the ATLAS experiment.
}
\label{fig:result}
\end{figure}

Figure~\ref{fig:result} shows the values of $v/v_s$ that can explain the measured value of the DM energy density, $\Omega h^2 = 0.12$, for a given parameter set.
The values of $g_{\chi \chi h} c_{\theta_h}/v$ are also shown.
There is a one-to-one correspondence between $v/v_s$ and $g_{\chi \chi h} c_{\theta_h}/v$.
A different value of $\theta_h$ requires a different value of $v/v_s$.
However, the required values of $g_{\chi \chi h} c_{\theta_h} /v$ are independent from the choice of $\theta_h$ because $g_{\chi \chi h} c_{\theta_h}/v$ is proportional to $g_{\chi \chi h} g_{f\bar{f} h}$, which is a combination of the relevant couplings for the DM annihilation processes.
The figure shows the large enhancement in the coupling compared to
the one determined without taking into account the effect of the early kinetic decoupling ($T_\chi = T$ in the figure).
This large enhancement is a consequence of the suppression in the scattering amplitude, which is shown in Eq.~\eqref{eq:scatt-amp}.
We also find almost no difference between the QCD-A and QCD-B for $m_\chi \gtrsim 58$~GeV.
This result is also a consequence of the highly suppressed elastic scattering amplitude.
We conclude that it is necessary to include the effect of the early kinetic decoupling in the Higgs resonant region in the pNG model.

\begin{figure}[tb]
\includegraphics[width=0.48\hsize]{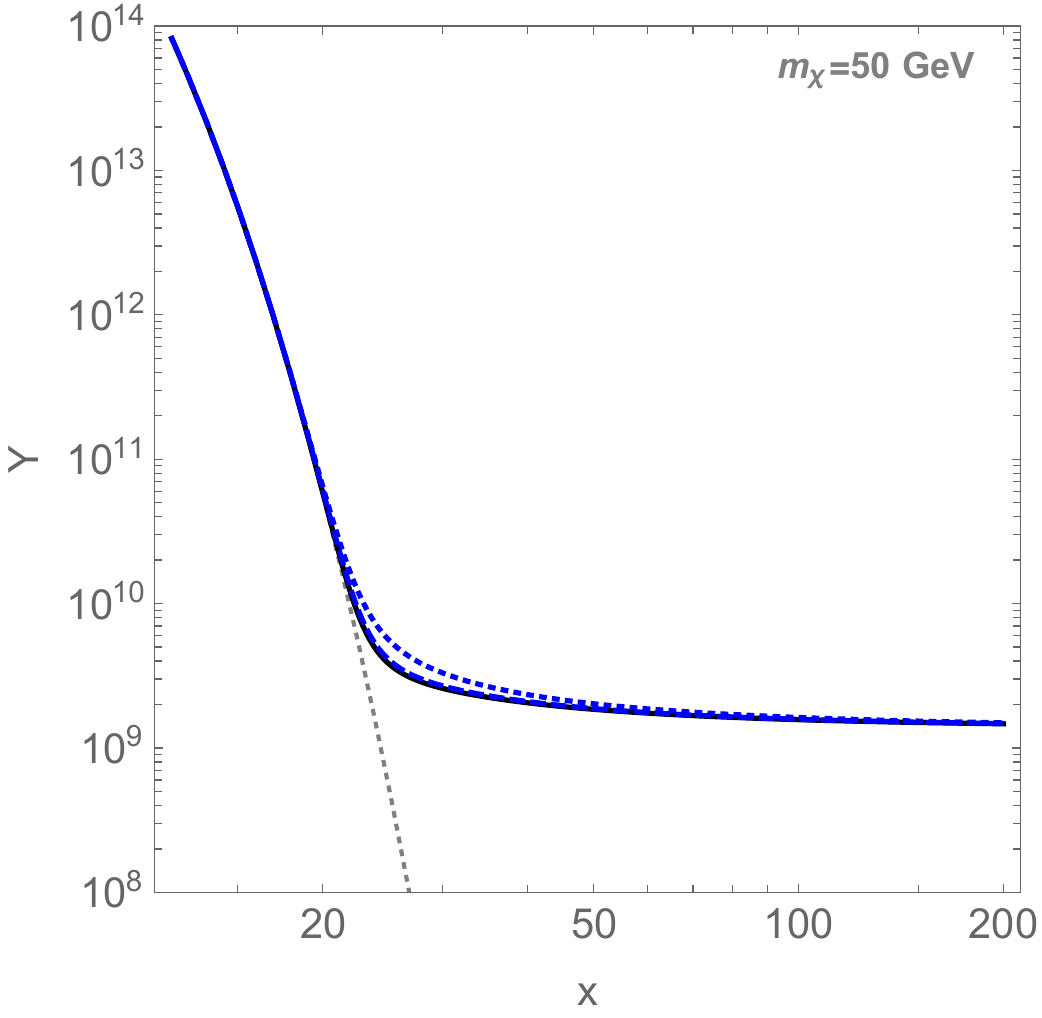}
\includegraphics[width=0.48\hsize]{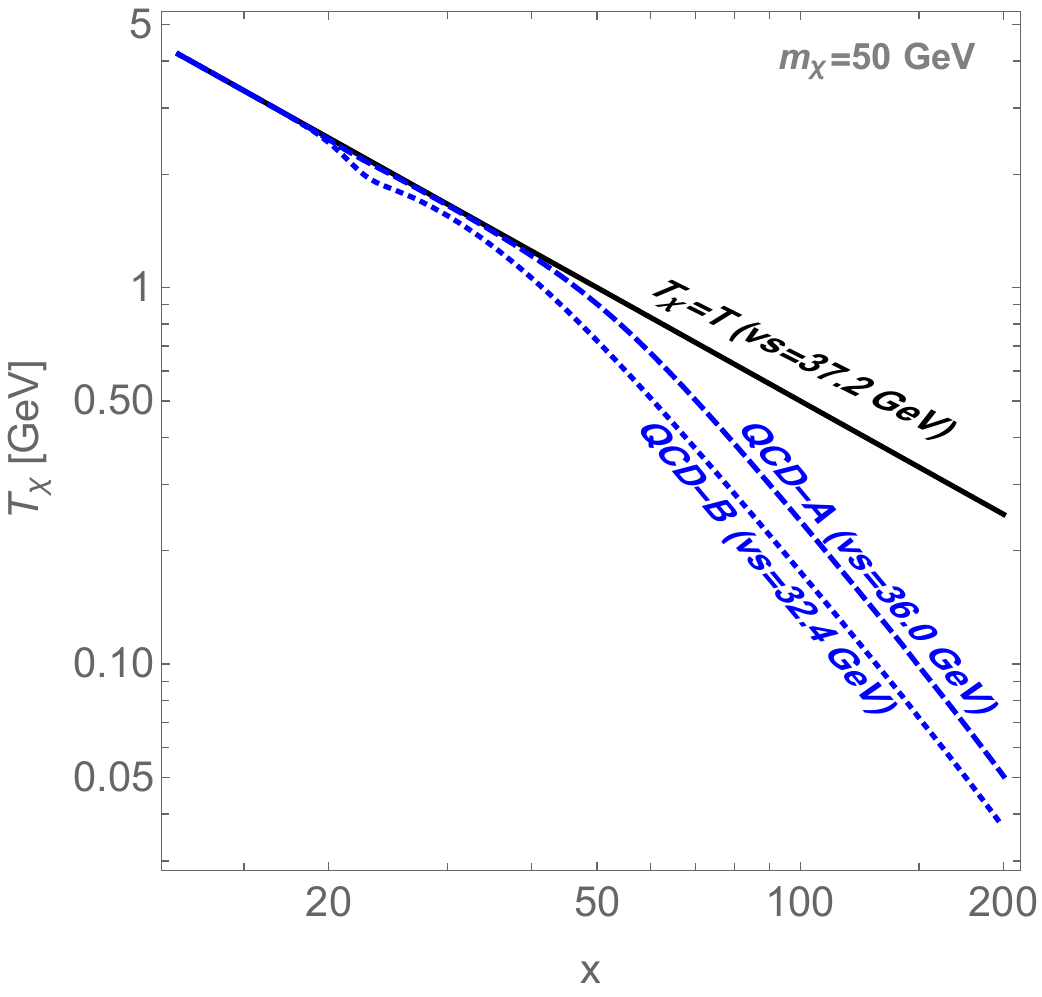}
\includegraphics[width=0.48\hsize]{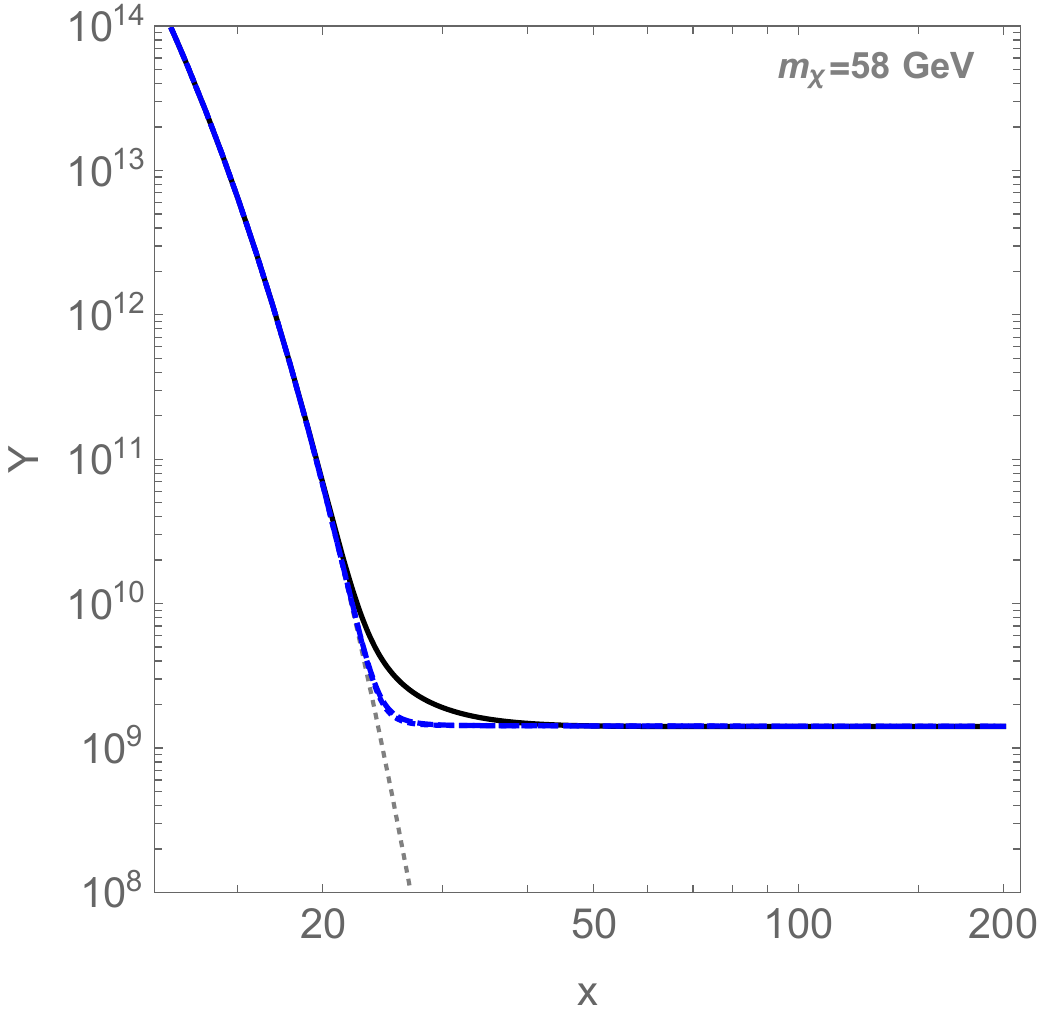}
\includegraphics[width=0.48\hsize]{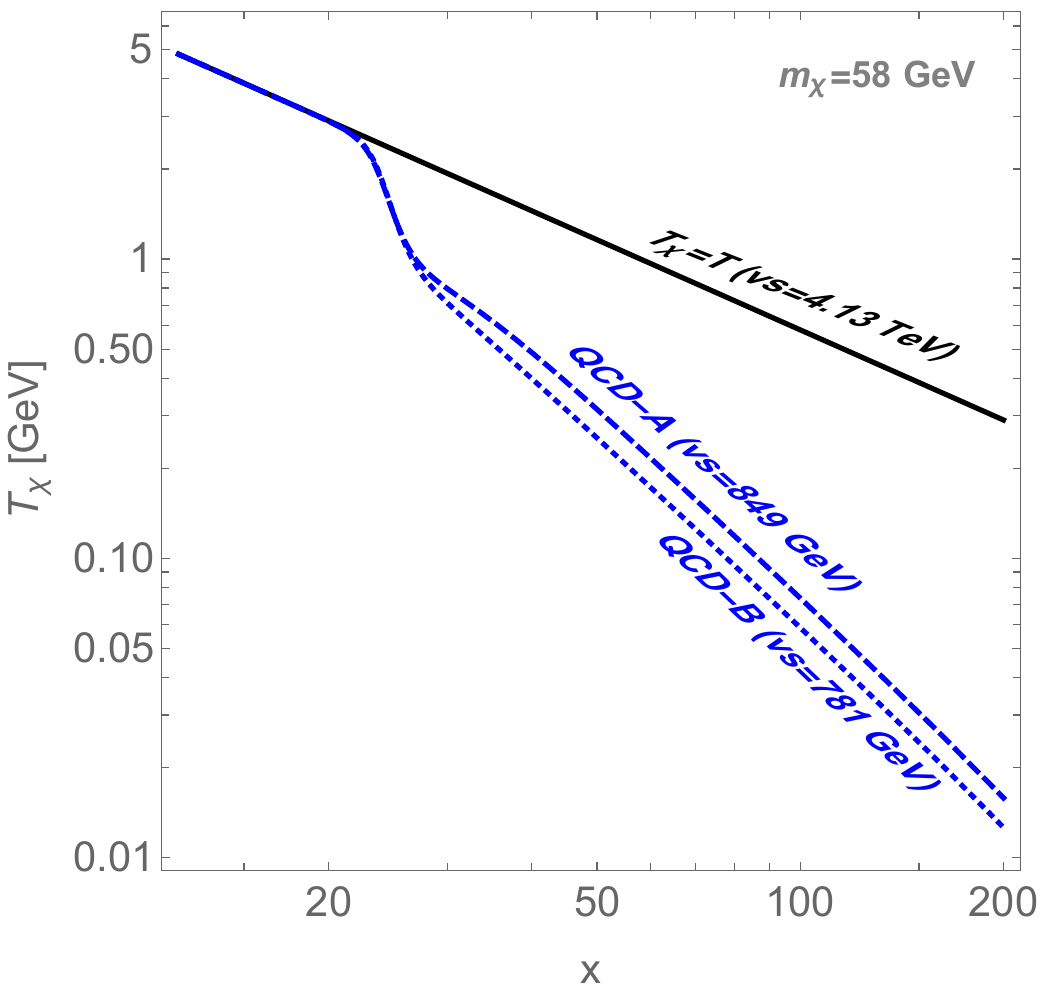}
\caption{
The evolution of $Y$ and $T_\chi$ for $m_\chi = $ 50~GeV (upper panels) and 58~GeV (lower panels).
The values of $v_s$ shown in the figure are determined to obtain the measured value of the DM energy density.
The black-solid, the blue-dashed, and the blue-dotted-dashed curves are for $T_\chi = T$,
 the QCD-A, and the QCD-B, respectively. In the left panels, the gray-dashed curve shows $Y_{eq}$.
}
\label{fig:temperature}
\end{figure}

%
%
%
In Fig.~\ref{fig:temperature}, we show the evolution of $Y$ and $T_\chi$ for $m_\chi = 50$~GeV and
58~GeV. Here $v_s$ is determined to obtain the measured value of the DM energy density as shown in Fig.~\ref{fig:result}. Therefore the left two panels, which show the evolution of $Y$, look almost the same.
From the evolution of $Y$, we can see that the freeze-out happens at $x \simeq 20$.
This behavior is very similar to the standard calculation where $T_\chi$ is assumed to be equal to $T$. 
As can be seen from the right panels, $T_\chi$ starts to differ from $T$ at $x \simeq 20$. 
Namely, the kinetic decoupling happens earlier than usual~\cite{1706.07433},
and thus the evolution of $T_\chi$ is important to determine the thermal relic abundance in this model.

\begin{figure}[tb]
\includegraphics[width=0.68\hsize]{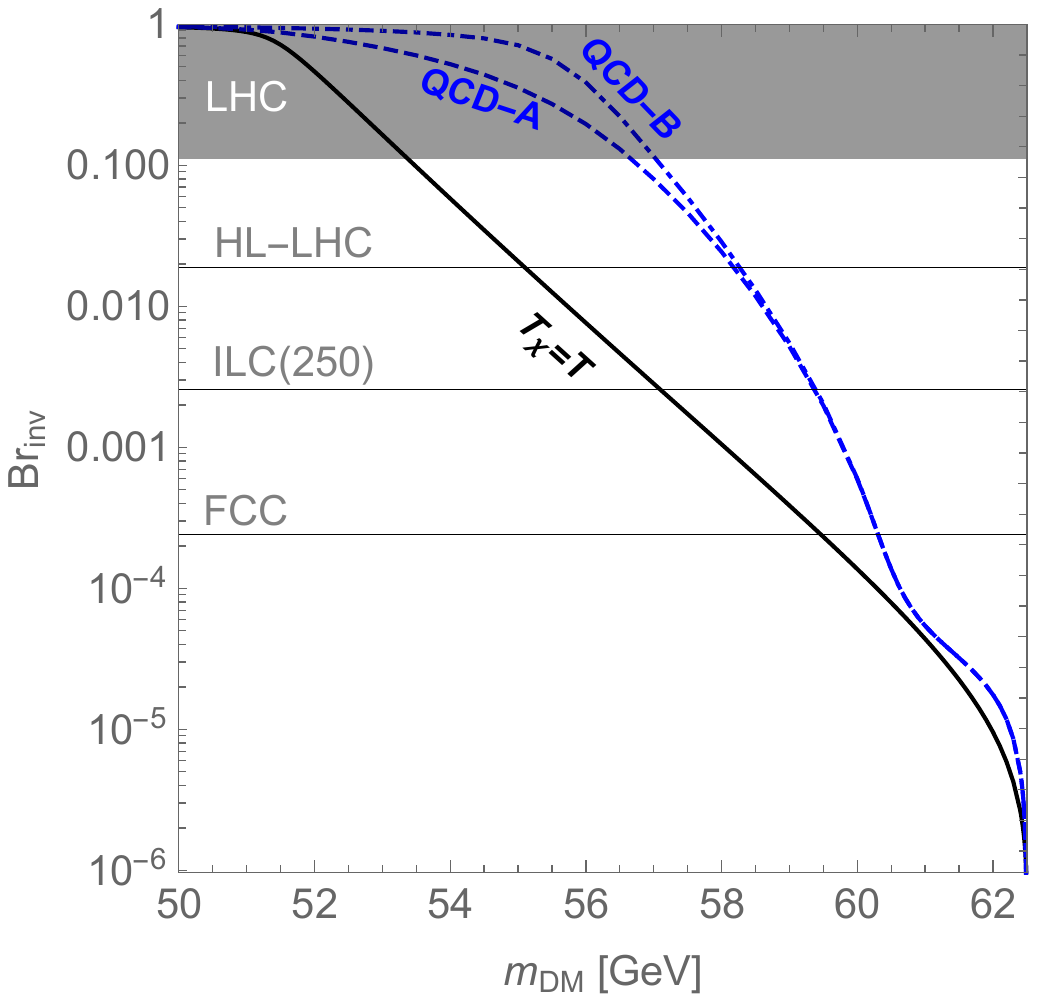}
\caption{
The values of the Higgs invisible decay.
The color notation is the same as in Fig.~\ref{fig:result}.
}
\label{fig:br}
\end{figure}

As an application of the early kinetic decoupling, 
we discuss the branching ratio of the SM Higgs boson into two DM particles. 
This process, known as the Higgs invisible decay, 
is enhanced by the enhancement in the DM-Higgs coupling
and is being searched by the ATLAS and CMS experiments. 
Currently, the ATLAS and CMS experiments obtain the upper bound on it as 
\begin{align}
 \text{BR}_\text{inv} <  
\begin{cases}
 0.11 & \text{(ATLAS \cite{ATLAS-CONF-2020-052})} \\
 0.19 & \text{(CMS \cite{1809.05937})} 
\end{cases}
\end{align}
at 95\% CL.
The prospects of various experiments are summarized in \cite{1905.03764},
\begin{align}
 \text{BR}_\text{inv} <
\begin{cases}
 0.019 & \text{(HL-LHC)} \\
 0.0026 & \text{(ILC(250))} \\
 0.00024 & \text{(FCC)}
\end{cases}
\end{align}
at 95\% CL,
where FCC corresponds to the combined performance of 
FCC-ee$_\text{240}$,
FCC-ee$_\text{365}$,
FCC-eh, and FCC-hh.
The prospects for the ILC and FCC are obtained by combining with the HL-LHC. 
In Fig.~\ref{fig:result}, 
we superimpose the current bound and the prospects of the Higgs invisible decay searches.
As can be seen, the effect of the early kinetic decoupling is significant.
The current lower mass bound on the DM is obtained as $\sim 57$~GeV, 
while it is about 53~GeV in the standard treatment where the effect of the kinetic decoupling is ignored. 
The prospects of the reach of the DM mass at future experiments also extended a few GeV.
We show the value of the Higgs invisible decay for a given DM mass in Fig.~\ref{fig:br}.
We find that difference in the predictions of the Higgs invisible decay with and without taking into account the effect of the early decoupling is as large as an order of magnitude for $m_\chi \lesssim 59$~GeV.

We emphasize that 
the significant enhancement in the coupling shown in Fig.~\ref{fig:result} is 
due to the suppression by the small momentum transfer in the scattering amplitude discussed in Eq.~\eqref{eq:scatt-amp}.
If the scattering amplitude is suppressed only by the small coupling and without small momentum transfer, 
the effect of the early kinetic decoupling is less efficient than in the pNG model.
For example, the scattering amplitude in the scalar singlet DM model~\cite{Silveira:1985rk, McDonald:1993ex, Burgess:2000yq} is suppressed only by the small coupling and is not suppressed by the small momentum transfer.
In that model, the coupling enhancement by the effect of the early kinetic decoupling~\cite{1706.07433, 1912.02870} is much milder than the one obtained in the pNG model.


\section{Conclusion}\label{sec:summary}
We have investigated the effect of the early kinetic decoupling in a pNG DM model.
We have focused on the Higgs resonant region, 50~GeV$\lesssim m_\chi \lesssim m_h/2$.
It is well known that the DM-Higgs coupling should be highly suppressed to obtain the measured value of the DM energy density by the freeze-out mechanism in that mass range. 
Moreover, thanks to the virtue of the pNG DM model, the DM-SM scattering processes are suppressed by low momentum transfer.
Therefore, the DM-SM scattering processes are suppressed both by the small coupling and the small momentum transfer. Thus the effect of the early kinetic decoupling is expected to be sizable.

We have shown that this suppression makes the effect of the early kinetic decoupling significant.
In order to obtain the measured value of the DM energy density by the freeze-out mechanism, the DM-Higgs coupling has to be larger than the value determined without taking into account the effect of the early kinetic decoupling.
An interesting consequence is the enhancement of the Higgs invisible decay.
As shown in Fig.~\ref{fig:br}, the Higgs invisible branching ratio is enhanced more than 
an order of magnitude in most of the region of the parameter space. 
This enlarges the discovery potential of DM at the collider experiments.

\section*{Acknowledgments}
This work is supported in part by JSPS KAKENHI Grant Numbers 19H04615 and 21K03549.
The author wishes to thank Tobias Binder for helpful discussions about the collision term.
The work is also supported by JSPS Core-to-Core Program (grant number:JPJSCCA20200002).

\appendix
\section*{Appendix}
\renewcommand{\thesubsection}{\Alph{section}.\arabic{section}}

\section{Collision term}
We discuss the DM elastic scattering off a particle in the thermal bath, and its contribution to the collision term.

We consider a two-to-two process, $\chi(1) f(3) \to \chi(2) f(4)$, where $\chi$ is DM, $f$ is a particle in the thermal bath, and the numbers are the indices for the momentum.
The contribution of this process to the collision term is given by
\begin{align}
{\cal C}_{el.}
=&
\frac{1}{g_\chi}\frac{1}{2}
\prod_{j = 2,3,4}\int \frac{d^3p_j}{(2\pi)^3 2 E_j}
\sum_{1,2,3,4}\left| {\cal M}_{13\to24}\right|^2
\left(
-
f_1 f^{eq.}_3 
\left( 1 \pm f_2   \right)
\left( 1 \pm f^{eq.}_4   \right)
+
f_2 f^{eq.}_4 
\left( 1 \pm f_1   \right)
\left( 1 \pm f^{eq.}_3   \right)
\right)
\nonumber\\
& 
\qquad \qquad \qquad \times
(2\pi)^4 \delta^{4}(p_1 + p_3 - p_2 - p_4)
.
\label{eq:c_el}
\end{align}
Using the following relation~\cite{1912.02870},
\begin{align}
 f^{eq.}_3 \left( 1 \pm f^{eq.}_4   \right)
=&
 \frac{1}{e^{E_3 \beta} \mp 1}  \left(1 \pm \frac{1}{e^{E_4 \beta} \mp 1} \right)
\nonumber\\
=&
 \frac{e^{E_4 \beta}}{\left( e^{E_3 \beta} \mp 1 \right) \left( e^{E_4 \beta} \mp 1\right)} 
\nonumber\\
=&
e^{(E_4-E_3) \beta}
 f^{eq.}_4 \left( 1 \pm f^{eq.}_3  \right)
,
\end{align}
we can simplify the collision term as
\begin{align}
{\cal C}_{el.}
=&
\frac{1}{g_\chi}\frac{1}{2}
\prod_{j = 2,3,4}\int \frac{d^3p_j}{(2\pi)^3 2 E_j}
\sum_{1,2,3,4}\left| {\cal M}_{13\to24}\right|^2
\left(
-
f_1 \left( 1 \pm f_2   \right)
+
f_2 \left( 1 \pm f_1   \right)
e^{(E_3-E_4) \beta}
\right)
\nonumber\\
& 
\qquad \qquad \qquad \times
 f^{eq.}_3 \left( 1 \pm f^{eq.}_4   \right)
(2\pi)^4 \delta^{4}(p_1 + p_3 - p_2 - p_4)
.
\end{align}
We further simplify this collision term with the following two assumptions:
\begin{itemize}
 \item The distribution of the DM is given by $f(E,T) = \alpha(T) e^{- E/T_\chi} = \frac{n_\chi}{n_\chi^{eq.}(T_\chi)} e^{-E/T_\chi}$.
 \item The scattering amplitude depends on $t$ but independent of $s$.
\end{itemize} 
The first assumption is justified if DM sufficiently interacts with particles in the dark sector even after the decoupling from the thermal bath. It is easily realized by introducing the DM self-interaction. The second assumption is realized in simple DM models such as the pNG DM model.

Under these assumptions,
we begin with the integration with respect to $p_4$,
\begin{align}
{\cal C}_{el.}
=&
\frac{1}{g_\chi}\frac{1}{2}
\prod_{j = 2,3}\int \frac{d^3p_j}{(2\pi)^3 2 E_j}
\frac{1}{2 (E_1 + E_3 - E_2)}
\sum_{1,2,3,4}\left| {\cal M}_{13\to24}\right|^2
\left(
-
f_1 \left( 1 \pm f_2   \right)
+
f_2 \left( 1 \pm f_1   \right)
e^{(E_2-E_1) \beta}
\right)
\nonumber\\
& 
\qquad \qquad \qquad \times
 f^{eq.}_3 \left( 1 \pm f^{eq.}_4   \right)
(2\pi) \delta(E_1 + E_3 - E_2 - \sqrt{(\vec{p_1}+\vec{p_4}-\vec{p_2})^2 + m_f^2 })
.
\end{align}
The argument of the squared root in the delta function becomes zero when
\begin{align}
 (\vec{p_1}+\vec{p_4}-\vec{p_2})^2 + m_f^2 
=&
 (\vec{p_1}-\vec{p_2})^2 +\vec{p_4}^2 + m_f^2 
+ 2 |\vec{p_1}-\vec{p_2}||\vec{p_4}| \cos\theta_*,
\end{align}
where $\theta_*$ is the angle between $\vec{p_4}$ and $\vec{p_1}- \vec{p_2}$.

Here we use the second assumption, namely $\sum_{1,2,3,4}|{\cal M}_{13 \to 24}|^2$ depends only on $t$ and independent of $s$.
Then the integral with respect to $\cos\theta_*$ does not affect to $\sum_{1,2,3,4}|{\cal M}_{13 \to 24}|^2$,
and the scattering term is simplified as
\begin{align}
{\cal C}_{el.}
=&
\frac{1}{g_\chi}\frac{1}{2}
\int \frac{d^3p_2}{(2\pi)^3 2 E_2}
\frac{1}{|\vec{p_1}-\vec{p_2}|}
\sum_{1,2,3,4}\left| {\cal M}_{13\to24}\right|^2
\left(
-
f_1 \left( 1 \pm f_2   \right)
+
f_2 \left( 1 \pm f_1   \right)
e^{(E_2-E_1) \beta}
\right)
\nonumber\\
& 
\qquad \qquad \qquad \times
\int \frac{dE_3}{8\pi}
 f^{eq.}_3 \left( 1 \pm f^{eq.}_4   \right)
\theta
\left(
1 - \frac{(t + 2 E_3 (E_1 - E_2))^2}{4 \vec{p_3}^2 (\vec{p_2}-\vec{p_1})^2}
\right)
\\
=&
\pm \frac{1}{\beta}
\frac{1}{g_\chi}
\frac{1}{256\pi^3}
\frac{1}{|\vec{p}_1|}
\int_m^{\infty} dE_2
\int_{t_{min.}}^{t_{max.}}  dt
\frac{1}{\sqrt{(E_2 - E_1)^2 - t}}
\noindent\\
& \qquad \qquad \qquad
\sum_{1,2,3,4}\left| {\cal M}_{13\to24}\right|^2
\left(
-
f_1 \left( 1 \pm f_2   \right)
+
f_2 \left( 1 \pm f_1   \right)
e^{(E_2-E_1) \beta}
\right)
\nonumber\\
& 
\qquad \qquad \qquad \times
\frac{1}{1 - e^{-\beta (E_1 - E_2)}} 
\ln
\left(
\frac{1 \mp e^{-\beta(E_3^* + E_1 - E_2) }
}{1 \mp e^{-\beta E_3^*  }
}
\right)
,
\end{align}
where
\begin{align}
 t_{max}=&
 (E_1 - E_2)^2 - (|\vec{p}_1| - |\vec{p}_2|)^2
,\\
 t_{min}=&
 (E_1 - E_2)^2 - (|\vec{p}_1| + |\vec{p}_2|)^2
,\\
 E_3^* 
 =& \frac{1}{2}
\left(
E_2 - E_1 
+ \sqrt{(E_2-E_1)^2 - t}\sqrt{1 - \frac{4 m_q^2}{t}}
\right).
\end{align}
Using $f(E,T) = \alpha(T) e^{- E/T_\chi} = \frac{n_\chi}{n_\chi^{eq.}(T_\chi)} e^{-E/T_\chi}$, 
we find
\begin{align}
-
f_1 \left( 1 \pm f_2   \right)
+
f_2 \left( 1 \pm f_1   \right)
e^{(E_2-E_1) \beta}
\simeq&
- f_1 + f_2 e^{(E_2-E_1) \beta}
\nonumber\\
=&
- \alpha(T) e^{- E_1/T_\chi}
\left(
1 - e^{(E_2-E_1) (\beta- \beta_\chi)}
\right),
\end{align}
where $\beta_\chi = T_\chi^{-1}$.

Finally, we find that
\begin{align}
& g_\chi \int \frac{d^3 p_1}{(2\pi)^3} \frac{1}{E_1} G(E_1)
{\cal C}_{el.}
\nonumber\\
=&
\pm \frac{1}{\beta}
\frac{1}{512\pi^5}
\alpha(T)
\int_m^\infty d E_1
\int_m^{\infty} dE_2
 G(E_1)
\int_{t_{min.}}^{t_{max.}}  dt
\frac{1}{\sqrt{(E_2 - E_1)^2 - t}}
\nonumber\\
& \qquad \qquad \qquad
\sum_{1,2,3,4}\left| {\cal M}_{13\to24}\right|^2
\left(
-  e^{- E_1 \beta_\chi}
\frac{1 - e^{(E_2-E_1) (\beta- \beta_\chi)}
}{
1 - e^{\beta (E_2 - E_1)} 
}\right),
\nonumber\\
& 
\qquad \qquad \qquad \times
\ln
\left(
\frac{1 \mp e^{-\beta(E_3^* + E_1 - E_2) }
}{1 \mp e^{-\beta E_3^*  }
}
\right)
,
\label{eq:full}
\end{align}
where $G(E_1)$ is an arbitrary function of $E_1$.

For $G(E_1) = 1$, Eq.~\eqref{eq:full} should vanish because the number density of DM does not change by the elastic scattering processes.
In Eq.~\eqref{eq:full}, $E_1$ and $E_2$ are dummy indices and can be renamed as $E_2$ and $E_1$, respectively. Namely, we can exchange $E_1$ and $E_2$. For $G(E_1) = 1$, the integrand flips its sign under the exchange of $E_1$ and $E_2$. 
Therefore Eq.~\eqref{eq:full} vanishes for $G(E_1) = 1$.

\section{The second moment}

We substitute $\vec{p}_1^2/E_1 = (E_1^2 - m^2)/E_1$ into $G(E_1)$ in Eq.~\eqref{eq:full}.
Note that the integrand other than $G(E)$ flips its sign under the exchange of $E_1$ and $E_2$.
In other words, the integrand is an odd function of $E_1 - E_2$.
Using this fact, we replace the integrand and the integral intervals as 
\begin{align}
 \frac{\vec{p_1}^2}{E_1}  \to& \frac{1}{2} \left( \frac{\vec{p_1}^2}{E_1} - \frac{\vec{p_2}^2}{E_2}   \right),\\
 \int_m^\infty dE_1  \int_m^\infty dE_2 \to&   2 \int_m^\infty dE_1  \int_m^{E_1} dE_2.
\end{align}
After this simplification, we find 
\begin{align}
& g_\chi \int \frac{d^3 p_1}{(2\pi)^3} \frac{1}{E_1} \frac{\vec{p_1}^2}{E_1}
{\cal C}_{el.}
\nonumber\\
=&
\pm \frac{1}{\beta}
\frac{1}{512\pi^5}
\alpha(T)
\int_m^\infty d E_1
\int_m^{E_1} dE_2
\left(
\frac{\vec{p_1}^2}{E_1} - \frac{\vec{p_2}^2}{E_2}
\right)
\left(
-  e^{- E_1 \beta_\chi}
\frac{1 - e^{(E_2-E_1) (\beta- \beta_\chi)}
}{
1 - e^{\beta (E_2 - E_1)} 
}\right)
\nonumber\\
& \qquad \qquad \qquad
\times 
\int_{t_{min.}}^{t_{max.}}  dt
\sum_{1,2,3,4}\left| {\cal M}_{13\to24}\right|^2
\frac{1}{\sqrt{(E_2 - E_1)^2 - t}}
\nonumber\\
& 
\qquad \qquad \qquad \times
\ln
\left(
\frac{1 \mp 
\exp\left(\beta \frac{E_2 - E_1}{2}\right)
\exp\left(-\frac{\beta}{2}
\sqrt{(E_2-E_1)^2 - t}
\sqrt{1 - \frac{4 m_q^2}{t}}
\right)
}{1 \mp 
\exp\left(-\beta \frac{E_2 - E_1}{2}\right)
\exp\left(-\frac{\beta}{2}
\sqrt{(E_2-E_1)^2 - t}
\sqrt{1 - \frac{4 m_q^2}{t}}
\right)
}
\right)
,
\label{eq:full_2nd}
\end{align}

For the numerical evaluation, it is better to change the variables from $(E_1, E_2, t)$ to $(y, z, \omega)$, where
\begin{align}
 y =& \frac{E_2 - m}{E_1 -m},\\
 z =& \frac{t - (2 m^2 - 2E_1 E_2 - 2 |\vec{p}_1| |\vec{p}_2|)}{4 |\vec{p}_1| |\vec{p}_2|},\\
 \omega =& \frac{E_1 - m}{E_1}.
\end{align}
We find
\begin{align}
& g_\chi \int \frac{d^3 p_1}{(2\pi)^3} \frac{1}{E_1} \frac{\vec{p_1}^2}{E_1}
{\cal C}_{el.}
\nonumber\\
=&
\pm \frac{1}{\beta}
\frac{\alpha(T)m^4}{128\pi^5}
\int_0^1 d \omega
\int_0^1 dy
\int_0^1 dz
\sqrt{\epsilon_1^2 - 1}
\sqrt{\epsilon_2^2 - 1}
\epsilon_1^2 ( \epsilon_1 - 1)
(\epsilon_1 - \epsilon_2)
\left(
1 + \frac{1}{\epsilon_1 \epsilon_2}
\right)
\nonumber\\
& \qquad \qquad \qquad \times
\frac{-e^{-\epsilon_1 x_\chi} + e^{(\epsilon_2 - \epsilon_1) x} e^{- \epsilon_2 x_\chi}}{ 1 - e^{x(\epsilon_2 - \epsilon_1)}}
\frac{1}{\sqrt{(\epsilon_1 - \epsilon_2)^2 - \frac{t}{m^2}}}
\sum_{1,2,3,4}\left| {\cal M}_{13\to24}\right|^2
\nonumber\\
& 
\qquad \qquad \qquad \times
\ln
\left(
\frac{1 \mp e^{-\frac{x(\epsilon_1 - \epsilon_2)}{2}} \exp\left(-\frac{x}{2}\sqrt{(\epsilon_1 - \epsilon_2)^2 - \frac{t}{m^2}}\sqrt{1 - \frac{4 m_f^2}{t}}\right)
}{
1 \mp e^{\frac{x(\epsilon_1 - \epsilon_2)}{2}}\exp\left(-\frac{x}{2}\sqrt{(\epsilon_1 - \epsilon_2)^2 - \frac{t}{m^2}}\sqrt{1 - \frac{4 m_f^2}{t}}\right)
}
\right)
,
\end{align}
where
\begin{align}
 \epsilon_1 =& \frac{E_1}{m} = \frac{1}{1-\omega},\\
 \epsilon_2 =& \frac{E_1}{m} = 1+ \frac{\omega y}{1-\omega},\\
 t =& m^2 
\left( 4 \sqrt{\epsilon_1^2 - 1} \sqrt{\epsilon_2^2 - 1} z + (\epsilon_1-\epsilon_2)^2 -( \sqrt{\epsilon_1^2 - 1} + \sqrt{\epsilon_2^2 - 1} )^2  \right).
\end{align}

The evolution of the DM temperature of $y$ depends on \cite{1706.07433}
\begin{align}
 \frac{1}{y} \frac{d y}{dx}
\supset&
\sqrt{\frac{8 m_{pl}^2 \pi^2}{45}} \frac{m}{x^2} \sqrt{g_*(T)} \frac{m}{s} 
 \frac{1}{3 n_\chi T_\chi m}
 g_\chi \int \frac{d^3 p_1}{(2\pi)^3} \frac{1}{E_1} \frac{\vec{p_1}^2}{E_1} {\cal C}_{el.}
\nonumber\\
\equiv&
\frac{\sqrt{g_*(T)}}{h(T)} 
\tilde{\delta}
.
\label{eq:dy/dx}
\end{align}
Here, we define $\tilde{\delta}$.\footnote{$\tilde{\delta}$ is related to $\tilde{\gamma}$ used in Ref.~\cite{1706.07433} 
$\tilde{\delta} = x^2 \left( \frac{y_{eq}}{y} - 1 \right) \tilde{\gamma}$
}
We find $\tilde{\delta}$ defined here is given by
\begin{align}
 \tilde{\delta}
=&
\frac{15}{g_\chi}
\sqrt{\frac{8 m_{pl}^2 \pi^2}{45}} 
 \frac{m}{T_\chi^2}
\frac{e^{-x_\chi}}{K_2(x_\chi)}
\left( \pm \frac{1}{128\pi^5} \right)
\int_0^1 d \omega
\int_0^1 dy
\int_0^1 dz
\nonumber\\
& \qquad \qquad \qquad \times
\sqrt{\epsilon_1^2 - 1}
\sqrt{\epsilon_2^2 - 1}
\epsilon_1^2 ( \epsilon_1 - 1)
(\epsilon_1 - \epsilon_2)
\left(
1 + \frac{1}{\epsilon_1 \epsilon_2}
\right)
\nonumber\\
& \qquad \qquad \qquad \times
\frac{-e^{-(\epsilon_1 -1)x_\chi} + e^{(\epsilon_2 - \epsilon_1) x} e^{- (\epsilon_2 -1)x_\chi}}{ 1 - e^{x(\epsilon_2 - \epsilon_1)}}
\frac{1}{\sqrt{(\epsilon_1 - \epsilon_2)^2 - \frac{t}{m^2}}}
\sum_{1,2,3,4}\left| {\cal M}_{13\to24}\right|^2
\nonumber\\
& 
\qquad \qquad \qquad \times
\ln
\left(
\frac{1 \mp e^{-\frac{x(\epsilon_1 - \epsilon_2)}{2}} \exp\left(-\frac{x}{2}\sqrt{(\epsilon_1 - \epsilon_2)^2 - \frac{t}{m^2}}\sqrt{1 - \frac{4 m_f^2}{t}}\right)
}{
1 \mp e^{\frac{x(\epsilon_1 - \epsilon_2)}{2}}\exp\left(-\frac{x}{2}\sqrt{(\epsilon_1 - \epsilon_2)^2 - \frac{t}{m^2}}\sqrt{1 - \frac{4 m_f^2}{t}}\right)
}
\right)
.
\end{align}
The upper (lower) sign is for the bosonic (fermionic) particle in the thermal bath.


\bibliography{refs}

\begin{thebibliography}{10}

\bibitem{1807.06209}
Planck, N.~Aghanim {\em et~al.},
\newblock Astron. Astrophys. {\bf 641}, A6 (2020), 1807.06209.

\bibitem{Lee:1977ua}
B.~W. Lee and S.~Weinberg,
\newblock Phys. Rev. Lett. {\bf 39}, 165 (1977).

\bibitem{1608.07648}
LUX, D.~S. Akerib {\em et~al.},
\newblock Phys. Rev. Lett. {\bf 118}, 021303 (2017), 1608.07648.

\bibitem{1708.06917}
PandaX-II, X.~Cui {\em et~al.},
\newblock Phys. Rev. Lett. {\bf 119}, 181302 (2017), 1708.06917.

\bibitem{1805.12562}
XENON, E.~Aprile {\em et~al.},
\newblock Phys. Rev. Lett. {\bf 121}, 111302 (2018), 1805.12562.

\bibitem{1203.2064}
L.~Lopez-Honorez, T.~Schwetz, and J.~Zupan,
\newblock Phys. Lett. {\bf B716}, 179 (2012), 1203.2064.

\bibitem{1404.3716}
S.~Ipek, D.~McKeen, and A.~E. Nelson,
\newblock Phys. Rev. {\bf D90}, 055021 (2014), 1404.3716.

\bibitem{1609.09079}
M.~Escudero, A.~Berlin, D.~Hooper, and M.-X. Lin,
\newblock JCAP {\bf 1612}, 029 (2016), 1609.09079.

\bibitem{1612.06462}
M.~Escudero, D.~Hooper, and S.~J. Witte,
\newblock JCAP {\bf 1702}, 038 (2017), 1612.06462.

\bibitem{1708.02253}
C.~Gross, O.~Lebedev, and T.~Toma,
\newblock Phys. Rev. Lett. {\bf 119}, 191801 (2017), 1708.02253.

\bibitem{1706.07433}
T.~Binder, T.~Bringmann, M.~Gustafsson, and A.~Hryczuk,
\newblock Phys. Rev. {\bf D96}, 115010 (2017), 1706.07433,
\newblock [Erratum: Phys. Rev.D101,no.9,099901(2020)].

\bibitem{1912.02870}
K.~Ala-Mattinen and K.~Kainulainen,
\newblock JCAP {\bf 2009}, 040 (2020), 1912.02870.

\bibitem{1901.08074}
A.~Hektor, A.~Hryczuk, and K.~Kannike,
\newblock JHEP {\bf 03}, 204 (2019), 1901.08074.

\bibitem{2004.10041}
T.~Abe,
\newblock Phys. Rev. {\bf D102}, 035018 (2020), 2004.10041.

\bibitem{2101.04887}
S.~Abe, G.-C. Cho, and K.~Mawatari,
\newblock (2021), 2101.04887.

\bibitem{2001.03954}
Y.~Abe, T.~Toma, and K.~Tsumura,
\newblock JHEP {\bf 05}, 057 (2020), 2001.03954.

\bibitem{2001.05910}
N.~Okada, D.~Raut, and Q.~Shafi,
\newblock Phys. Rev. {\bf D103}, 055024 (2021), 2001.05910.

\bibitem{2104.13523}
Y.~Abe, T.~Toma, K.~Tsumura, and N.~Yamatsu,
\newblock (2021), 2104.13523.

\bibitem{2105.03419}
N.~Okada, D.~Raut, Q.~Shafi, and A.~Thapa,
\newblock (2021), 2105.03419.

\bibitem{2103.01944}
T.~Binder, T.~Bringmann, M.~Gustafsson, and A.~Hryczuk,
\newblock (2021), 2103.01944.

\bibitem{1307.1347}
LHC Higgs Cross Section Working Group, J.~R. Andersen {\em et~al.},
\newblock (2013), 1307.1347.

\bibitem{1503.03513}
M.~Drees, F.~Hajkarim, and E.~R. Schmitz,
\newblock JCAP {\bf 1506}, 025 (2015), 1503.03513.

\bibitem{Tanabashi:2018oca}
Particle Data Group, M.~Tanabashi {\em et~al.},
\newblock Phys. Rev. {\bf D98}, 030001 (2018).

\bibitem{1205.1914}
P.~Gondolo, J.~Hisano, and K.~Kadota,
\newblock Phys. Rev. {\bf D86}, 083523 (2012), 1205.1914.

\bibitem{0903.0189}
T.~Bringmann,
\newblock New J. Phys. {\bf 11}, 105027 (2009), 0903.0189.

\bibitem{ATLAS-CONF-2020-052}
ATLAS, T.~A. collaboration,
\newblock (2020).

\bibitem{1809.05937}
CMS, A.~M. Sirunyan {\em et~al.},
\newblock Phys. Lett. {\bf B793}, 520 (2019), 1809.05937.

\bibitem{1905.03764}
J.~de~Blas {\em et~al.},
\newblock JHEP {\bf 01}, 139 (2020), 1905.03764.

\bibitem{Silveira:1985rk}
V.~Silveira and A.~Zee,
\newblock Phys. Lett. {\bf 161B}, 136 (1985).

\bibitem{McDonald:1993ex}
J.~McDonald,
\newblock Phys. Rev. {\bf D50}, 3637 (1994), hep-ph/0702143.

\bibitem{Burgess:2000yq}
C.~P. Burgess, M.~Pospelov, and T.~ter Veldhuis,
\newblock Nucl. Phys. {\bf B619}, 709 (2001), hep-ph/0011335.

\end{thebibliography}
\bibliographystyle{h-physrev} 
\end{document}